\documentclass[12pt]{iopart}
\usepackage[british]{babel}
\usepackage[utf8]{inputenc}
\usepackage{amsmath, amssymb, bm}
\usepackage{graphicx}
\usepackage{appendix}
\usepackage{hyperref}

\renewcommand{\vec}[1]{\mathbf{#1}}

\begin{document}

\title[Exact Diagonalisation of phBECs]{Exact Diagonalisation of Photon Bose-Einstein Condensates\\with Thermo-Optic Interaction}

\author{Enrico Stein}
\ead{estein@rhrk.uni-kl.de}
\address{Department of Physics and Research Center OPTIMAS, Technische Universität Kaiserslautern, Erwin-Schrödinger Straße 46, 67663 Kaiserslautern, Germany}

\author{Axel Pelster}
\ead{axel.pelster@physik.uni-kl.de}
\address{Department of Physics and Research Center OPTIMAS, Technische Universität Kaiserslautern, Erwin-Schrödinger Straße 46, 67663 Kaiserslautern, Germany}

\vspace{10pt}
\begin{indented}
\item[]\today
\end{indented}

\begin{abstract}
Although photon Bose-Einstein condensates have already been used for studying many interesting effects, the precise role of the photon-photon interaction is not fully clarified up to now. In view of this, it is advantageous that these systems allow measuring both the intensity of the light leaking out of the cavity and its spectrum at the same time. Therefore, the photon-photon interaction strength can be determined once via analysing the condensate broadening and once via examining the interaction-induced modifications of the cavity modes. As the former method depends crucially on the concrete shape of the trapping potential and the spatial resolution of the used camera, interferometric methods promise more precise measurements.\\
To this end, the present paper works out the impact of the photon-photon interaction upon the cavity modes. A quantum mechanical description of the photon-photon interaction, including the thermal cloud, builds the theoretical backbone of the method. An exact diagonalisation approach introduced here exposes how the effective photon-photon interaction modifies both the spectrum and the width of the photon gas. A comparison with a variational approach based on the Gross-Pitaevskii equation quantifies the contribution of the thermal cloud in the respective applications.
\end{abstract}

%
\vspace{2pc}
%
\noindent{\submitto{\NJP}}

\maketitle

\section{Introduction}
\begin{figure}
    \centering
    \includegraphics[width=.75\linewidth]{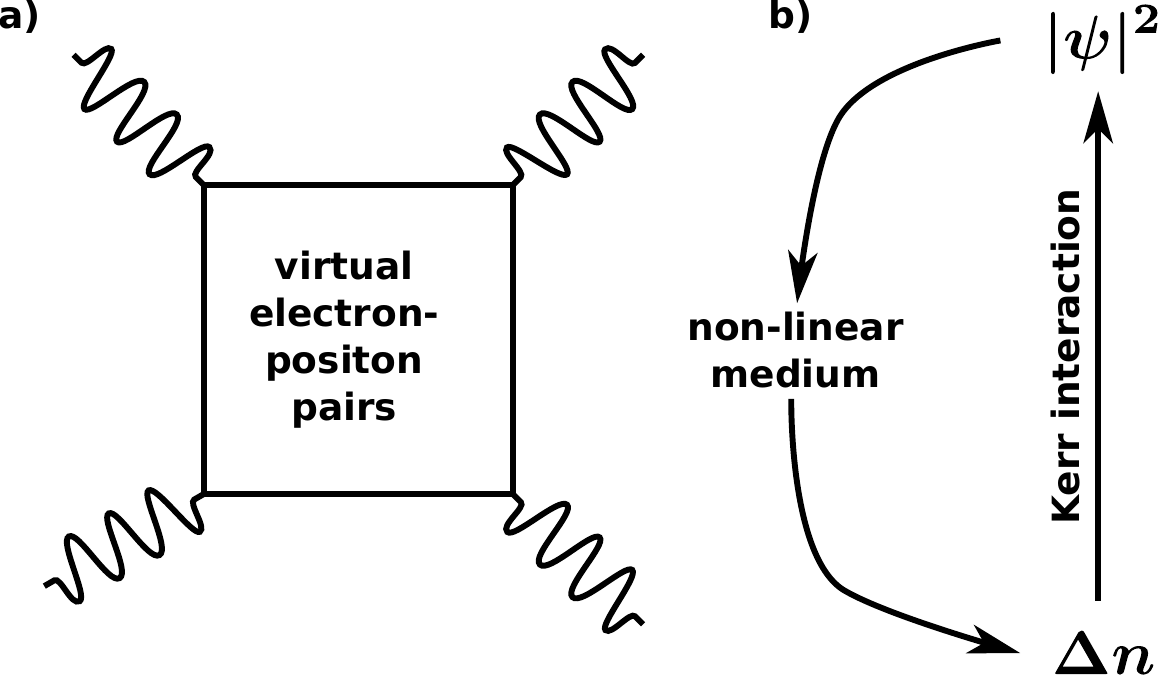}
    \caption{Possible photon-photon interaction processes. \textbf{a)} Interaction vertex in vacuum. Two incoming photons produce a virtual electron-positron pair, which recombines again. \textbf{b)} Photons in non-linear media, here represented via their density $|\psi|^2$, interact with the non-linear medium. This effectively introduces a Kerr interaction between two photons via changes  of the refractive index $\Delta n$.}
    \label{Fig:interactions_a}
\end{figure}

Observing photon-photon interactions is a demanding task. Classical electrodynamics considers light as a linear phenomenon, such that two crossing light beams only interfere with each other, but do not scatter. However, the advent of quantum electrodynamics changed this view, since in this theory electromagnetic fields can polarise the vacuum. Halpern was the first to express the idea of light-by-light scattering within the Dirac theory of electrons and positrons \cite{Halpern1933}. Afterwards, references \cite{Euler1935} and \cite{Euler1936} formalised the idea already to the modern picture, as figure \ref{Fig:interactions_a} a) portrays it. Here, two photons interact by polarising the vacuum via the production of a virtual electron-positron pair. The pair then recombines back into two photons with different wave vectors. Subsequently, reference \cite{Heisenberg1936} works out the corresponding modification of the Maxwell equations in vacuum, which is nowadays referred to as the Euler-Heisenberg Lagrangian. However, the authors of reference \cite{Heisenberg1936} point out, that this only works with photons in the Röntgen- or gamma-ray regime, since the energy needs to be high enough for supporting the electron-positron pair creation. The later works \cite{Karplus1950, Karplus1951} support these findings by applying the more modern $S$-matrix apparatus. Since this process appears in fourth-order perturbation theory, the cross-section for this kind of photon-photon interaction is proportional to the fourth power of Sommerfeld's constant. Therefore, only particle accelerators allow access to this kind of processes. In 2017 the ATLAS experiment at the Large Hadron Collider was able to observe light-by-light scattering in vacuum for the first time \cite{ATLAS2017}.\\
Embedding photons in non-linear materials increases the photon-photon
interaction significantly. One prominent example for such a non-linear process is the Kerr effect \cite{Boyd2008}, see figure \ref{Fig:interactions_a} b). Here, the refractive index of the material changes with the photon density, leading to effects like the self-focusing of a light beam or the existence of optical solitons. Such experiments are performed in effectively two-dimensional setups. This is remnant from light propagation in, e.g., glass cylinders, where the coordinate along the optical axis, due to the paraxial approximation, acts as the time coordinate and the remaining two dimensions as true spatial dimensions \cite{Lax1975, Zangwill}. A second possibility relies on confining the light together with the non-linear medium inside a cavity. As the cavity mirrors impose Dirichlet boundary conditions upon the light field, a standing wave emerges along the optical axis, freezing out the motion along this direction. However, the used photons are usually prepared in a dissipative state, as light leaks out of the apparatus and has to be reinjected by an external light source for achieving a steady state.\\
However, filling a microcavity with a dye solution offers the possibility to
\begin{figure}
    \centering
    \includegraphics[width=\linewidth]{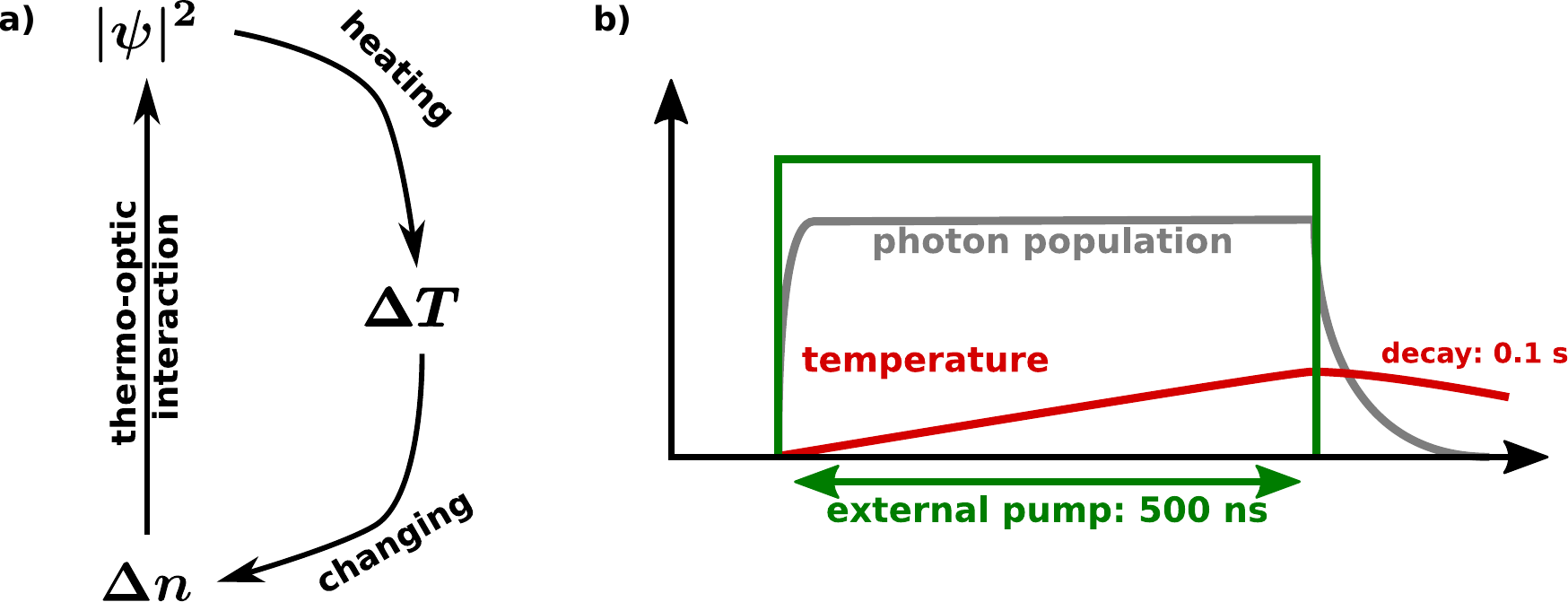}
    \caption{Photon-photon interaction in photon BEC setups. \textbf{a)} Mechanism of thermo-optic interaction. The photon density $|\psi|^2$ heats the medium to a temperature $\Delta T$. This shifts the refractive index by $\Delta n$, which effectively introduces a photon-photon interaction. \textbf{b)} Timescales of thermo-optic interaction. The timescales of the photon population and the external pump pulse are the fastest timescales, whereas the scale of the temperature diffusion is the slowest one.}
    \label{Fig:interactions_b}
\end{figure}
create light, the steady-state spectral distribution of which resembles thermal equilibrium \cite{Klaers2010, Klaers2011}. In these systems, also a small effective photon-photon interaction emerges, which does not stem from a Kerr effect but from a thermal non-linearity, the mechanism of which figure \ref{Fig:interactions_b} a) sketches. This originates from a heating of the dye molecules, which do not emit back all absorbed photons into the photon gas, but instead convert them into vibronic excitations. As a consequence, the dye solution heats up, changing its refractive index. As the heat diffuses through the experimental setup, the resulting effective thermo-optic photon-photon interaction is non-local in space and retarded in time, which represents the main difference to the photon-photon interactions sketched in figure \ref{Fig:interactions_a}. Since the temporal retardation is large compared to the remaining experimental timescales, see figure \ref{Fig:interactions_b} b), the thermo-optic photon-photon interaction effectively corresponds to a potential, which increases linearly in time \cite{Dung2017, Stein2022c}. As the thermo-optic interaction turns out to be weak, only the photon BEC (phBEC) regime allows determining its strength by measuring the condensate width \cite{Klaers2010}. However, due to the smallness of the effective photon-photon interaction, this method lacks of precision.\\
Instead, spectroscopic measurements promise an increased precision by measuring the small energy shifts, which the thermo-optic photon-photon interaction induces during a single pump pulse. The basis of the theoretical description of this thermo-optic interaction is a detailed model for the diffusion of the temperature produced in a single pump pulse. As the temperature slowly builds up, an adiabatic approximation of the corresponding quantum mechanical modelling grants access to the instantaneous eigenvalues. This allows determining the effective photon-photon interaction strength from the energy difference between two eigenmodes. Whilst reference \cite{Stein2022c} works out the corresponding basic theory and applies first-order perturbation theory for deriving some initial results, this paper goes beyond and works out in detail both the energy shifts and the condensate broadening by applying exact diagonalisation (ED) \cite{Weisse2008}.\\
The ED of the underlying second-quantised Hamiltonian represents a powerful method in case of increased effective photon-photon interactions. For instance, reference \cite{Stein2022b} predicts this situation at the dimensional crossover from 2D to 1D, where a significant trap anisotropy enhances the photon-photon interaction via an increased photon density. Experimentally, microstructured mirrors provide such anisotropic traps \cite{Maruo97, Deubel2004, Hohmann2015}. However, the thermal cloud leads to a stronger increase of the condensate width and to a possible overestimation of the effective photon-photon interaction, provided the theoretical modelling only considers the phBEC ground state. To avoid this, the ED method combines the investigation of the thermodynamic behaviour of the non-interacting phBEC \cite{Stein2022a} with the influence of the thermo-optic interaction upon the phBEC ground state at the dimensional crossover \cite{Stein2022b}.\\
The present paper is structured as follows: Section \ref{sec:ED} starts with a short summary of the underlying second-quantised Hamiltonian and uses ED for working out the new eigenmodes of the harmonically trapped photon gas. Moreover, the corresponding condensate width is compared to former results and special attention lies on the temperature dependence of both the eigenenergies and the width of the photon gas. Subsequently, section \ref{sec:Crossover} deals with the dimensional crossover and presents the corresponding results. Section \ref{sec:Summary} summarises the main findings of the paper.
\section{Exact Diagonalisation of Harmonic Potential}
\label{sec:ED}
This section provides a concise introduction to the model used for describing the thermo-optic interaction, which is based on reference \cite{Dung2017} and worked out more rigorously in the preceding paper \cite{Stein2022c}. Afterwards, ED serves as a method for benchmarking both the perturbative results on the energy spectrum derived in \cite{Stein2022c} and the variational approach performed in \cite{Dung2017}. The last part of this section goes beyond these findings and investigates the impact of finite temperatures on both the energy spectrum and the cloud width.

\subsection{Model}
As the phBEC itself varies on timescales, which are much faster than the produced temperature, see figure \ref{Fig:interactions_b} b),  reference \cite{Stein2022c} breaks down the quantum mechanical description of the thermo-optic interaction to an effective potential, which increases linearly in time. Hence, the second-quantised Hamiltonian,
\begin{align}\label{eq:Ham}
\hat H (t) = \sum_{\bm{n} \bm{n'}}\mathcal{H}_{\bm{n}, \bm{n'}} (t) \hat a^\dagger_{\bm n}(t) \hat a_{\bm{n'}}(t)\,,
\end{align}
bears an adiabatic time dependency introduced by the temporal retardation of the thermo-optic photon-photon interaction. The bosonic creation and annihilation operators $\hat a^\dagger_{\bm n}(t)$, $\hat a_{\bm{n'}}(t)$ belong to the instantaneous eigenmodes of the Hamiltonian \eqref{eq:Ham}. The Hamiltonian matrix in \eqref{eq:Ham} has the form
\begin{align}\label{eq:Hmat}
\mathcal H_{\bm n, \bm{n'}} (t)= E_{\bm n}(0) \delta_{\bm n, \bm{n'}} + g(t) F_{\bm n, \bm{n'}}\,,
\end{align}
with the effective time-dependent thermo-optic interaction strength $g(t)$ increasing linearly in time \cite{Stein2022c}. At the beginning of the experiment, i.e., at $t=0$, no interaction is present, yielding the initial value $g(0)=0$. The corresponding eigenvalue problem leads to the eigenenergies $E_{\bm n}(0)$ and the eigenfunctions $\psi_{\bm{n}}(\vec x)$. Furthermore, the non-diagonal matrix
\begin{align}\label{eq:Fmat}
F_{\bm n, \bm{n'}} = \sum_{\bm l} N_{\bm l}(0) \int d^2x~\psi_{\bm n}^* (\vec x) |\psi_{\bm l}(\vec x)|^2 \psi_{\bm{n'}}(\vec x)
\end{align}
contains the information about the thermo-optic interaction. It uses the mode occupation in the form of a Bose-Einstein distribution at the beginning of the experiment, which stems from the temporal retardation of the thermo-optic interaction:
\begin{align}
    N_{\bm l}(0) = \left\{ e^{\beta[E_{\bm l}(0) - \mu(0)] } - 1 \right\}^{-1} \, .
\end{align}
Here $\beta=1/(k_\text{B}T)$ denotes the inverse temperature and $\mu(0)$ represents the initial chemical potential, which is fixed by the conserved total particle number $N=\sum_{\bm l}\langle a^\dagger_{\bm l}(t) a_{\bm l}(t) \rangle$.
\subsection{Harmonic Potential}
This paper specialises to a harmonic trapping potential of the form 
\begin{align}\label{eq:Vho}
    V(\vec x) = \frac{m\Omega_x^2}{2}\,\left( x^2+\lambda^4 y^2\right),
\end{align}
with the effective photon mass $m$ and the trapping frequency $\Omega_x$ in $x$-direction. The trap-aspect ratio $\lambda = \sqrt{\Omega_y/\Omega_x}$ determines the trapping frequency $\Omega_y$ in $y$-direction. Therefore, the eigenenergies take the form
\begin{align}\label{eq:E0ho}
E_{\bm n}(0) = \hbar\Omega_x\left[ n_x + \lambda^2 n_y + (1+\lambda^2)/2 \right]\,.
\end{align}
Furthermore, the Gauß-Hermite functions
\begin{align}\label{eq:Psiho}
	\psi_{\bm n}(\vec x) = \sqrt{\frac{1}{2^{n_x+n_y}n_x!n_y!\pi l_x l_y}} \,H_{n_x}\left(\frac{x}{l_x}\right)H_{n_y}\left(\frac{y}{l_y}\right)e^{-x^2/(2l_x^2)-y^2/(2l_y^2)}
\end{align}
determine the eigenfunctions of the second-quantised Hamiltonian \eqref{eq:Ham} and the corresponding Hamiltonian matrix \eqref{eq:Hmat}, where $l_i = \sqrt{\hbar/(m\Omega_i)}$,  $i=x,y$, denote the oscillator lengths in both directions and $H_n$ are the Hermite polynomials. \\
This paper numerically determines the new eigenmodes of the Hamiltonian matrix \eqref{eq:Hmat} with the potential \eqref{eq:Vho} by applying exact diagonalisation. As this method necessitates using a finite number of Gauß-Hermite eigenmodes \eqref{eq:Psiho}, the sixty lowest energy subspaces are included. This ensures in the Bose-Einstein condensed regime, that the relative occupation of higher excited states is negligible. With this finite number of modes, the interaction matrix \eqref{eq:Fmat} is constructed for a given total particle number $N$ and thermodynamic temperature $T$. Hence, the presented method not only verifies previously published calculations relying on different methods, but also reveals possible deviations from the latter. To be specific, this section focuses on the isotropic case, i.e., $\lambda = 1$.
\subsection{Spectrum}
\begin{figure}
	\centering
	\includegraphics[width=\linewidth]{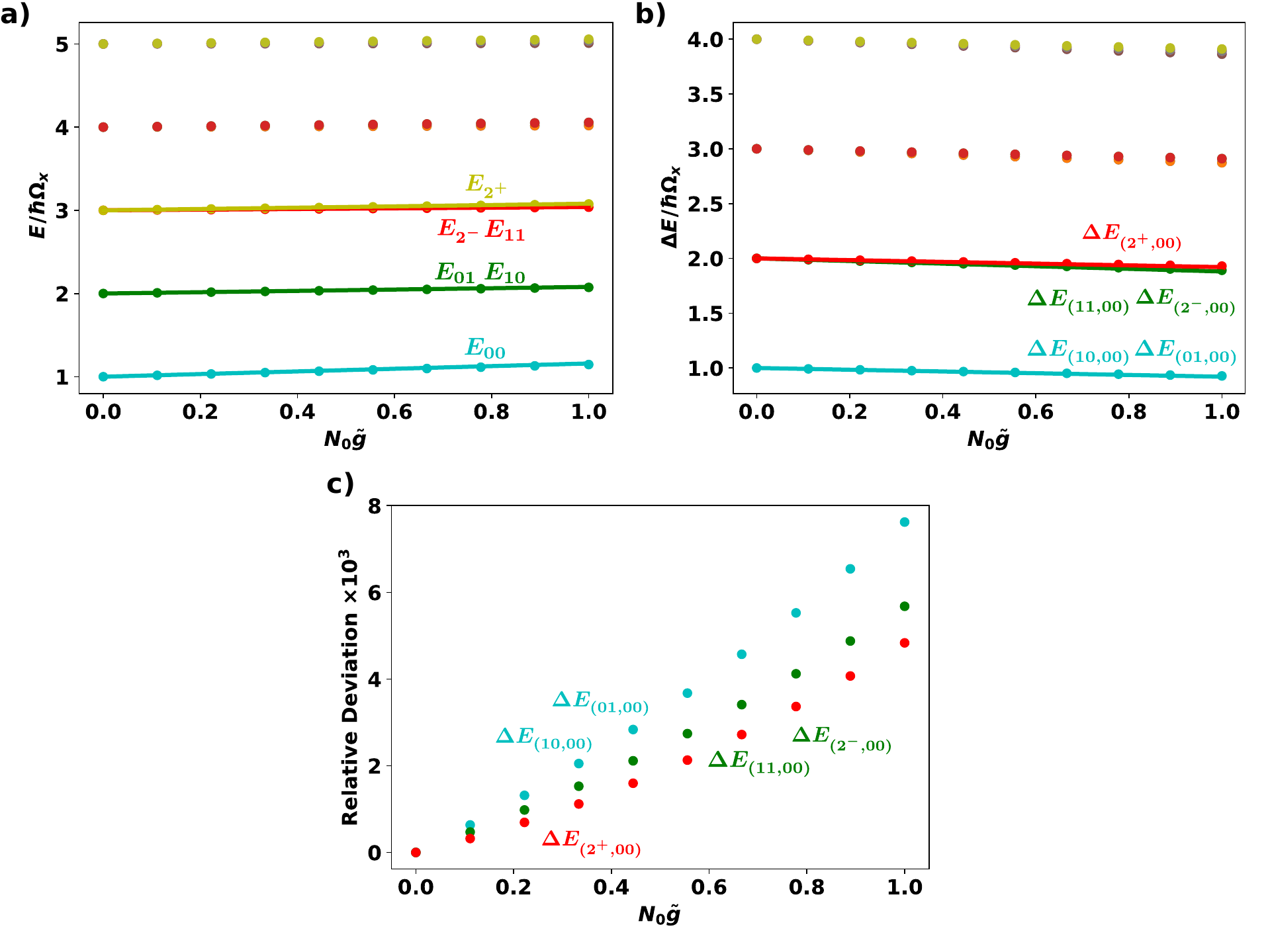}
	\caption{\textbf{a)} First few eigenenergies of the second-quantised Hamiltonian matrix \eqref{eq:Hmat}. The dots stand for the results of the ED, whereas the lines represent the first-order perturbative results from reference \cite{Stein2022c}. \textbf{b)} Corresponding energy differences $\Delta E_{\bm n, \bm{n'}} = E_{\bm n}-E_{\bm{n'}}$. \textbf{c)} Relative deviation of energy differences once calculated by the perturbation theory and once by the ED method. In all plots, the indices $2^\pm = [(20)\pm(02)]/\sqrt{2}$ denote the mode hybridisation due to the interaction.}
	\label{Fig:energy}
\end{figure}

This paragraph presents the results for the adiabatic time-dependent energy eigenvalues $E_{\bm n}(t)$ of the Hamilton matrix \eqref{eq:Hmat} for the potential \eqref{eq:Vho}. Figure \ref{Fig:energy} a) compares the eigenenergies obtained by the ED method with the results of first-order perturbation theory from the preceding paper \cite{Stein2022c}. It plots the eigenenergies as a function of the dimensionless interaction strength $\tilde g(t) = mg(t)/\hbar^2$, the magnitude of which is determined by the duration of the external pump beam. The results of both approaches agree well and, thus, the same observation holds for the corresponding energy differences shown in figure \ref{Fig:energy} b). However, for larger interaction strengths, a slight deviation of up to $10^{-3}$ between the ED results and the first-order perturbation theory \cite{Stein2022c} becomes visible in figure \ref{Fig:energy} c).
Hence, the perturbation theory derived in \cite{Stein2022c} yields a good approximation in this regime of parameters. Therefore, the corresponding formulas, which analytically express the photon-photon interaction strength via the energy differences, are precise enough for being applied to spectroscopic measurements.
\subsection{Condensate Width}
\begin{figure}
	\centering
	\includegraphics[width=.65\linewidth]{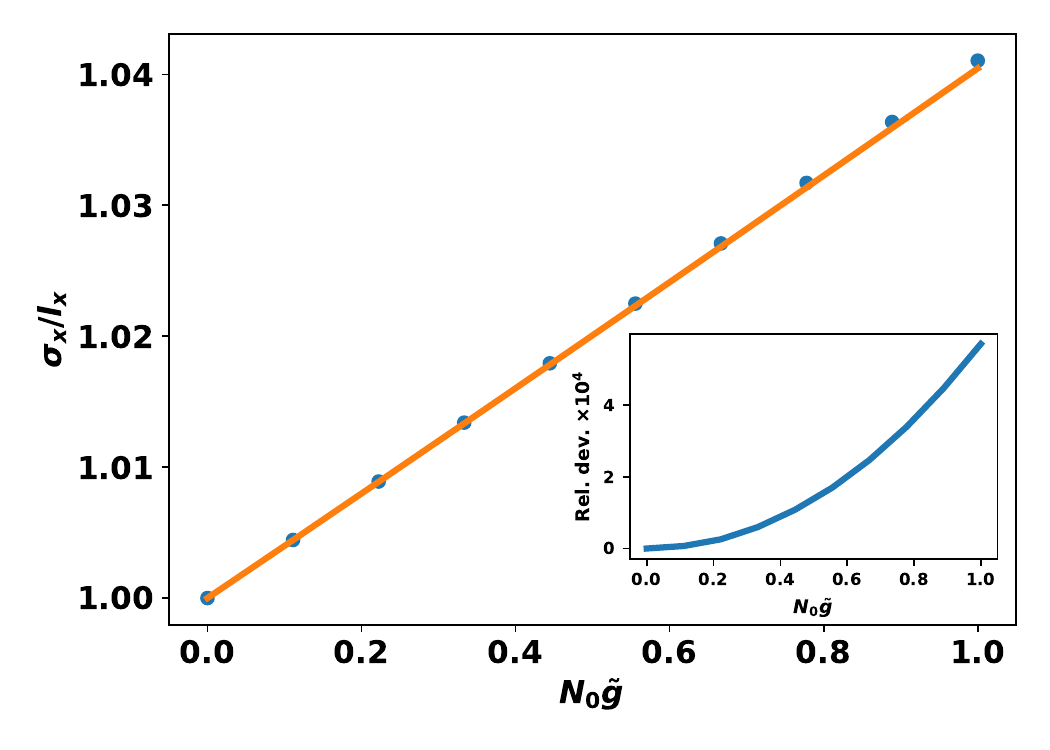}
	\caption{Comparison of the resulting condensate width $\sigma_x$ from the ED of the Hamiltonian matrix \eqref{eq:Hmat} (dots) with the variational approach from appendix \ref{App:Variational} (solid line). The inset shows the relative deviation between both results.}
	\label{Fig:Width_comp}
\end{figure}
The new eigenstates also show a broadening of the photon gas. The unitary matrix $U_{\bm n, \bm m}(t)$, which rotates into the instantaneous eigenbasis of the Hamiltonian matrix \eqref{eq:Hmat}, defines creation and annihilation operators $\hat a_{\bm n}(t), \, \hat a^\dagger_{\bm n}(t)$ at a given time $t$:
\begin{align}
    \hat a_{\bm n}(t) = \sum_{\bm l}U_{\bm n, \bm l}(t) \hat a_{\bm l}(0), \hspace*{1cm}\hat a^\dagger_{\bm n}(t) = \sum_{\bm l}U^\dagger_{\bm n, \bm l}(t) \hat a^\dagger_{\bm l}(0).
\end{align}
With these at hand, the total photon density,
\begin{align}\label{eq:density}
    n(\vec x, t) = \sum_{\bm l} N_{\bm l}(t) n_{\bm l}(\vec x, t),
\end{align}
is expressed in terms of the instantaneous eigenstate occupations,
\begin{align}\label{eq:Bose-Einstein}
    N_{\bm l}(t) = \left\{ e^{\beta[E_{\bm l}(t) - \mu(t)] } - 1 \right\}^{-1}\,,
\end{align}
and the corresponding eigenstate density,
\begin{align}
    n_{\bm l}(\vec x, t) = \sum_{\bm n, \bm m} U_{\bm l, \bm n}^\dagger(t) \psi_{\bm n}^*(\vec x)\psi_{\bm m}(\vec x)U_{\bm m, \bm l}(t) \, .
\end{align}
Therefore, the total width $\sigma_x$ of the photon gas is defined by the FWHM 
\begin{align}\label{eq:total_width}
	\sigma_x(t) = \sqrt{2\left\langle x^2 \right\rangle(t)}\,.
\end{align}
Here, the second moment of the density distribution
\begin{align}\label{eq:width_0}
    \left\langle x^2\right\rangle(t) = \frac{\int d^2x~x^2 n(\vec x,t)}{\int d^2x~n(\vec x,t)}\,
\end{align}
with the adiabatic time-dependent photon density \eqref{eq:density}, takes the form
\begin{align}
	\left\langle x^2\right\rangle (t) = \sum_{\bm l} \frac{N_{\bm{l}}(t)}{N}\, \left\langle x^2\right\rangle_{\bm{l}}(t) \,,
\end{align}
with the adiabatic time-dependent matrix elements
\begin{align}\label{eq:width_mat}
	\left\langle x^2\right\rangle_{\bm{l}}(t) = \int d^2x~x^2n_{\bm l}(\vec x, t)\,.
\end{align}
Older calculations and measurements \cite{Klaers2010, Dung2017} consider the deep condensate limit $N_{\bm 0}(t)\approx N$. In this case, \eqref{eq:total_width} reduces to the width of the ground state, i.e., $\sigma_x(t) \approx\sqrt{2\left\langle x^2 \right\rangle_{\bm 0}(t)}$. \ref{App:Variational} reproduces the corresponding variational calculation of the photon-con\-den\-sate width, which was first given in reference \cite{Dung2017}. Figure \ref{Fig:Width_comp} compares the results for the condensate width, calculated via ED and via the variational approach from \ref{App:Variational}. As the results agree well, the ED method developed here is found to be able to reliably reproduce the former theoretical and experimental results.\\
However, a closer look at the inset in figure \ref{Fig:Width_comp} for larger effective interaction strengths reveals that the results tend to deviate for larger effective photon-photon interaction strengths. Since the width calculated from the ED is larger than the corresponding width from the variational approach, this deviation stems from an increased coupling to the thermal cloud. Thus, the larger the photon-photon interaction strength is, the ground-state-only description is less accurate.

\subsection{Thermal Cloud}
\begin{figure}
	\centering
	\includegraphics[width=\linewidth]{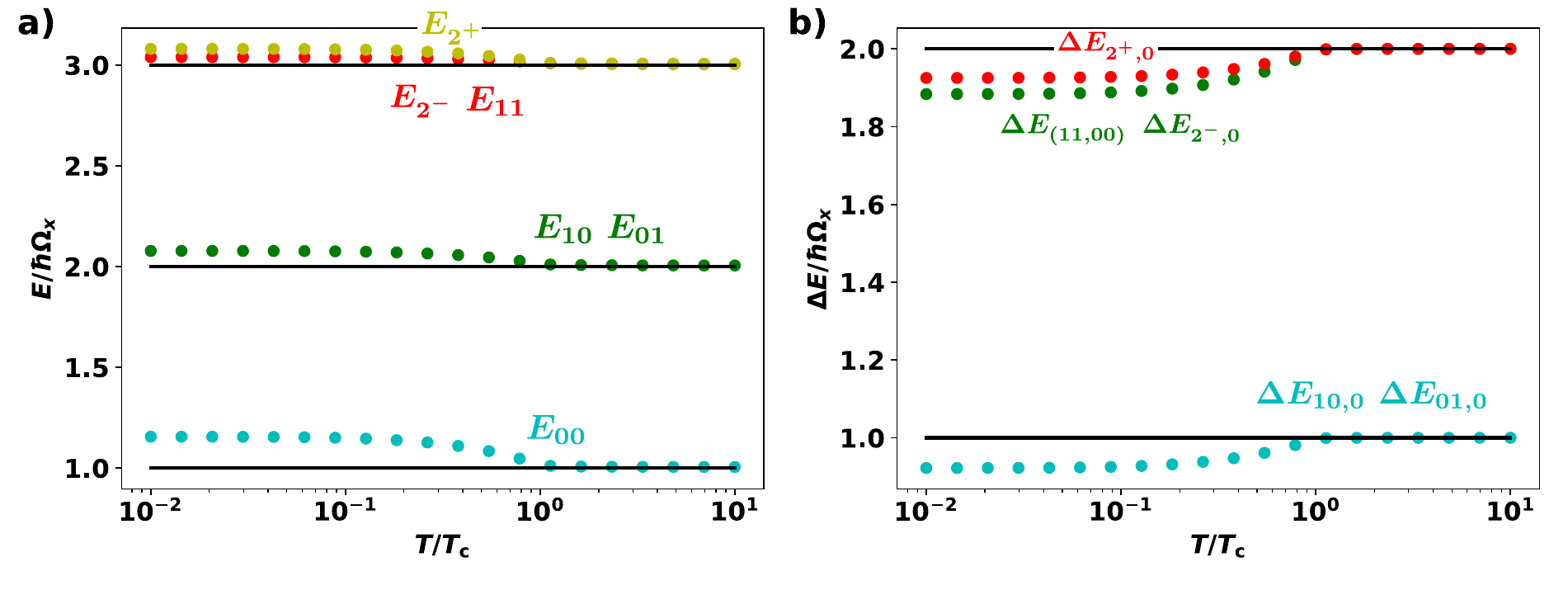}
	\caption{\textbf{a)} Temperature dependence of the first few energy eigenvalues of the Hamiltonian matrix \eqref{eq:Hmat}. \textbf{b)} Corresponding energy differences. The dots denote the results from the ED and the black lines are the corresponding \textbf{a)} energy eigenvalues and \textbf{b)} energy differences without the interaction, i.e., at the beginning of the experiment. Both plots are for total particle number $N = 10^4$ and interaction strength $\tilde g = 10^{-4}$.}
	\label{Fig:temp_energy}
\end{figure}
\begin{figure}
	\centering
	\includegraphics[width=.65\linewidth]{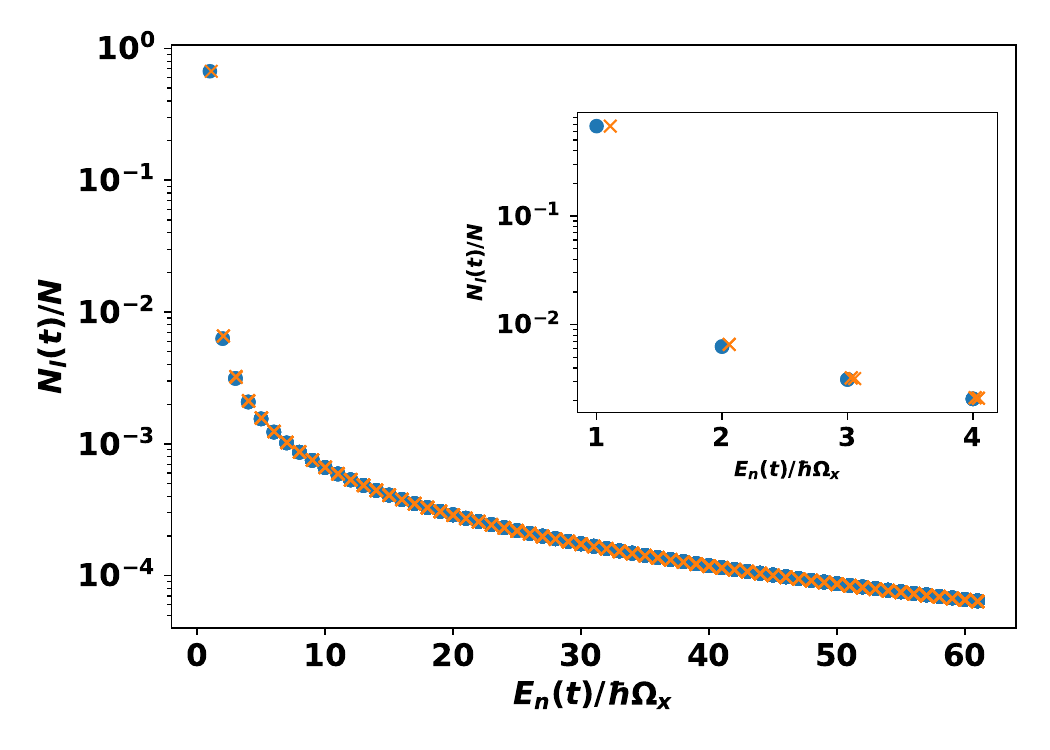}
	\caption{Photon occupation $N_{\bm l}(t)$ of the different states relative to the total photon number $N = 10^4$ in the Bose-Einstein condensed regime characterised by the condensate fraction $N_{\bm 0}/N\approx 0.9$. The dots (crosses) indicate the photon occupation without (with) interaction at the beginning (end) of the experiment. The inset shows the details for the lowest-lying states. The maximal interaction strength is $\tilde g(t_\text{end}) = 10^{-4}$.}
	\label{Fig:Distribution}
\end{figure}
\begin{figure}[t]
	\centering
	\includegraphics[width=.75\linewidth]{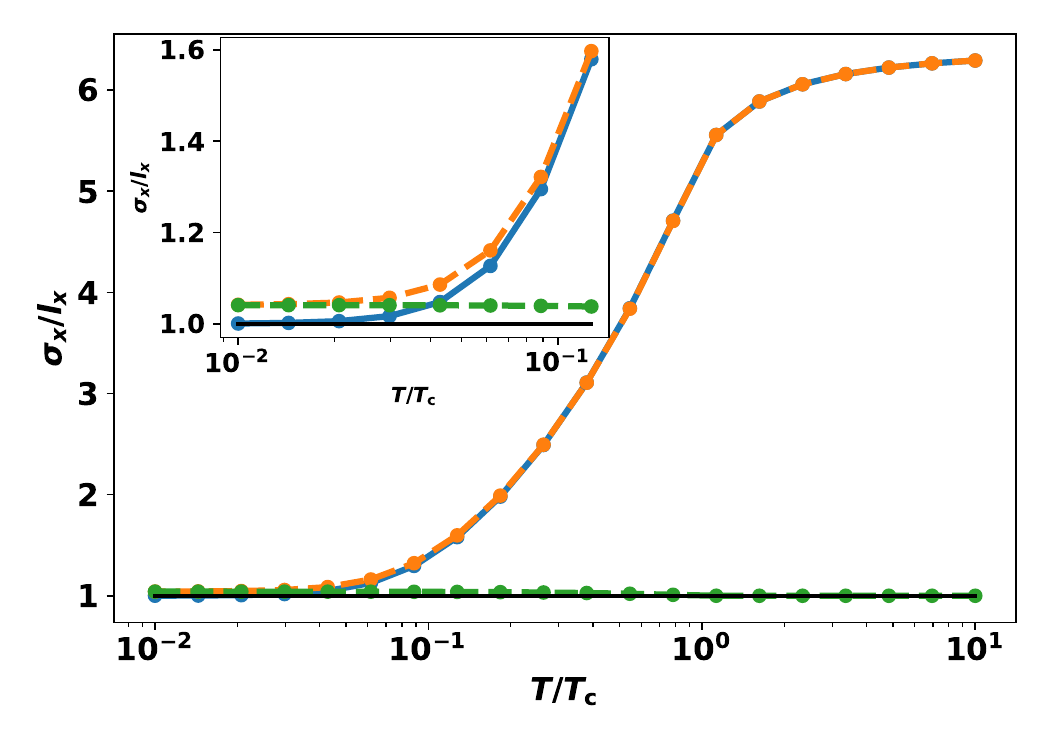}
	\caption{Total width of the photon gas at the beginning of the experiment (blue) and at the end of the experiment (orange), as well as the width of the ground state with interaction (green). The black line indicates the oscillator length, which is also the width of the ground state without interaction. The drawn lines are guides to the eye. The calculation uses the total particle number $N = 10^4$ and the maximal interaction strength $\tilde g(t_\text{end}) = 10^{-4}$.}
	\label{Fig:width_temp}
\end{figure}
Previous investigations of the effective photon-photon interaction only consider the zero-temperature limit. This section, however, deals with the effects of finite temperature, which the Hamiltonian matrix \eqref{eq:Hmat} naturally includes. At the beginning of the experiment, the thermo-optic interaction is not present, and the phBEC behaves as an ideal Bose gas in 2D \cite{Stein2022a}. As the photon-photon interaction strength increases along the duration of an experiment, this section aims at describing the influence of the thermal cloud upon the phBEC properties.\\
First, the focus lies on the temperature dependence of the energy eigenvalues, which figure \ref{Fig:temp_energy} a) shows. In the thermal regime, the energies undergo a small shift due to the interaction, whereas in the condensed regime, i.e., for $T<T_\text{c}$, a more significant shift builds up. This is due to the small ground-state occupation in the thermal phase. Since the coupling to the ground state is always the largest, significant energy shifts only occur in the condensed regime, as figure \ref{Fig:temp_energy} a) clearly indicates. Experiments can validate this finding by measuring the differences between the energy levels, which are plotted in figure \ref{Fig:temp_energy} b).\\
The time-dependent shifts of the energy eigenstates $E_{\bm l}(t)$ also affect the Bose-Einstein distribution \eqref{eq:Bose-Einstein} of the photons. Figure \ref{Fig:Distribution} compares the photon number distribution \eqref{eq:Bose-Einstein} at the beginning (dots) and at the end of the experiment (crosses). The horizontal shift of the occupation numbers at the end of the experiment indicates interaction-induced shifts of the eigenenergies. The occupation numbers themselves, however, undergo only minor modifications due to the smallness of the interaction.\\
Finally, the ED also allows calculating the width \eqref{eq:total_width} of the total photon gas, which figure \ref{Fig:width_temp} depicts. In the zero-temperature limit, the width of the photon gas approaches the ground-state width. At the beginning of the experiment, where no interaction is present, this width corresponds to the oscillator length, whereas during an experimental run the ground-state width increases and so the width of the total gas increases with the photon-photon interaction. In the thermal regime, however, the interaction does not lead to a noticeable change of the photon-gas width. This finding coincides with the observations above, showing only very small shifts in the energy levels in the thermal regime. The saturation of the width, which occurs for large temperatures, is a remnant from the finite number of considered eigenmodes.
\section{Dimensional Crossover}
\label{sec:Crossover}
With the development of new experimental techniques for micro-structuring the cavity mirrors \cite{Maruo97, Deubel2004, Hohmann2015}, also large trap anisotropies for phBECs can be manufactured. Therefore, these techniques allow realising a dimensional crossover from 2D to 1D. In this case, the ED method turns out to be essential, since reference \cite{Stein2022b} predicts a considerable enhancement of the effective photon-photon interaction strength.\\
The beginning of the section maps the 2D gas in a strong anisotropic trap onto a 1D gas, which determines the effective 1D interaction strength. The latter turns out to depend on the trap-aspect ratio $\lambda$. With this at hand, the effective 1D energy spectrum is calculated on the one hand numerically and on the other hand approximated analytically. The latter grants access to an approximation formula determining the effective photon-photon interaction strength from the differences of the eigenenergies in the quasi-1D limit. Finally, the calculation of the condensate width in the quasi-1D limit shows the significance of the ED method, as the width turns out to deviate from the variational approach presented in \cite{Dung2017}, which is recovered in \ref{App:Variational}. Note that in this chapter the final photon-photon interaction strength $\tilde g(t_\text{end})$ is assumed to be an order of magnitude smaller than in the previous section, such that $\tilde g(t_\text{end}) = 10^{-5}$. This adjustment takes into account the assumed shorter condensate lifetime due to the increased photon intensity, resulting from the strong anisotropy. Nevertheless, this section shows the strong increase of the effective photon-photon interaction strength due to the dimensional crossover. 

\subsection{Effective 1D Interaction Strength}
Reference \cite{Stein2022a} demonstrates that for trap-aspect ratios larger than a critical value $\lambda_\text{1D}$ the squeezed direction freezes out in its ground state and that the systems behaves effectively quasi one-dimensional. In addition, the density in the frozen out direction alters the bare 2D interaction strength $\tilde g$ \cite{Stein2022b}. The resulting effective quasi one-dimensional interaction strength $\tilde g_\text{1D}(\lambda)$ turns out to be a function of the trap-aspect ratio $\lambda$.\\
The current paper uses the same reasoning for a single pump pulse, which reference \cite{Stein2022b} applies to the steady state after several pump pulses. According to definition \eqref{eq:Vho} an increasing trap-aspect ratio squeezes the $y$-direction and, so, in the quasi-1D case the photons only populate the ground mode in this direction. This suggests the factorisation 
\begin{align}
\psi_{\bm n}(\vec x) = \chi_{n_x}(x) \varphi_{0}(y; \lambda)
\end{align}
for the Gauß-Hermite eigenmodes of the potential \eqref{eq:Vho}. Therefore, the interaction matrix \eqref{eq:Fmat} reduces to
\begin{align}\label{eq:F1D}
    F_{n_x, n'_x} = w(\lambda) \sum_{l_x} N_{(l_x, 0)} \int dx~\chi_{n_x}^*(x)|\chi_{l_x}(x)|^2\chi_{n'_x}(x)\,,
\end{align}
where the weight
\begin{align} \label{eq:1Dweight}
    w(\lambda) = \int dy~|\varphi_{0}(y;\lambda)|^4
\end{align}
incorporates the influence of the squeezed direction and, thus, depends on the trap-aspect ratio $\lambda$. In the real 1D case, however, the structure of the Hamiltonian matrix \eqref{eq:Hmat} remains the same; only the multi-indices turn into single indices. Hence, the identification
\begin{align}
    \tilde g_\text{1D}(t; \lambda) =  \tilde g(t) w(\lambda)
\end{align}
of an effective 1D interaction strength allows a mapping of the matrix \eqref{eq:Fmat} onto the real 1D case. The explicit calculation of the weight \eqref{eq:1Dweight} for the harmonic potential \eqref{eq:Vho} leads to
\begin{align}\label{eq:g1D}
    \tilde g_\text{1D}(t; \lambda) = \frac{\tilde g(t) \lambda}{\sqrt{2\pi}}\,.
\end{align}
The effective 1D interaction \eqref{eq:g1D} agrees with the corresponding result for the contact Kerr interaction obtained in reference \cite{Stein2022b}, but not with the similar one for the thermo-optic interaction, although the current work considers only the latter. The reason is the used short-time approximation in \eqref{eq:Ham}. Due to this, the matrix elements \eqref{eq:Fmat} effectively resemble the corresponding ones for a contact interaction. Reference \cite{Stein2022b} investigates the steady state after several pump pulses, such that the temperature diffusion has a significant impact on the behaviour. In the current work, however, this does not happen due to the short-time approximation.

\subsection{Energy Spectrum}
\begin{figure}
	\centering
	\includegraphics[width=\linewidth]{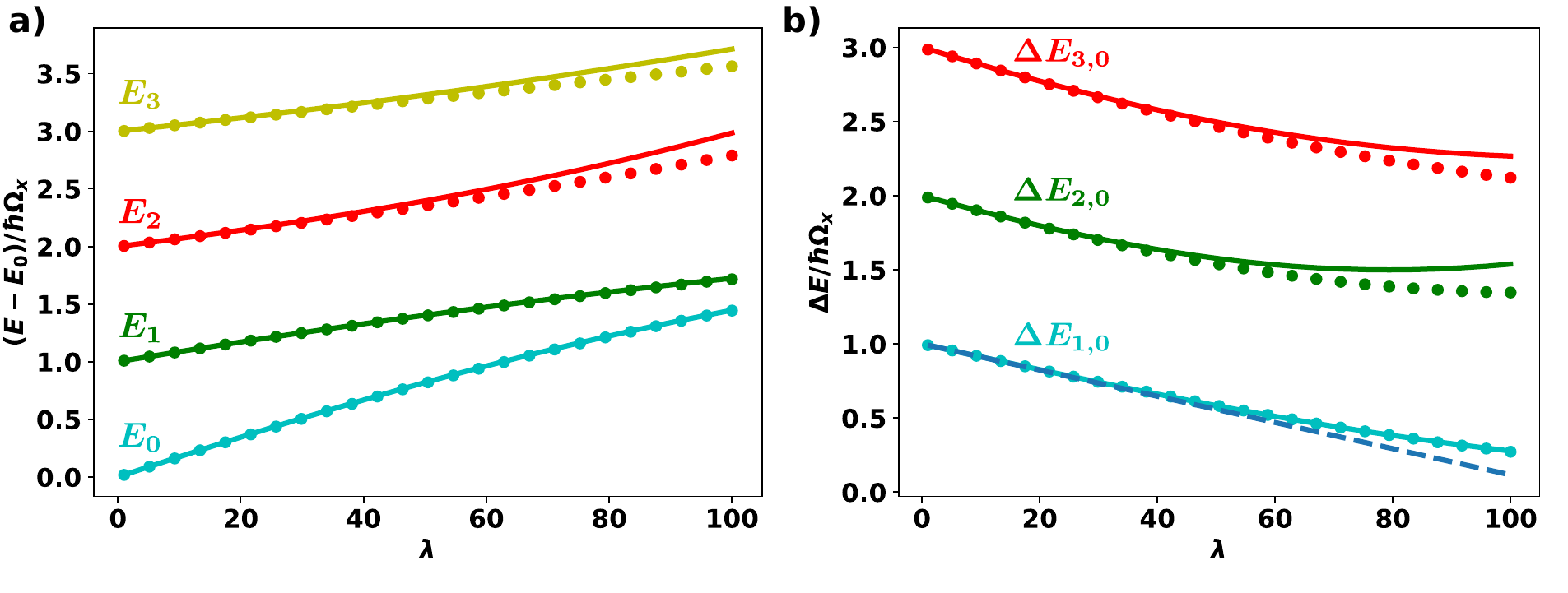}
	\caption{\textbf{a)} First few eigenenergies of the Hamiltonian matrix \eqref{eq:Hmat} relative to the ground state energy $E_0 = \hbar\Omega_x(1+\lambda^2)/2$ at dimensional crossover. \textbf{b)} Corresponding energy differences. Here, the dashed line shows the linearisation for the lowest energy difference $\Delta E_{1,0}$. In both pictures, the dots represent the results from the ED and the lines are the analytical approximations \eqref{eq:E1D}. The calculation uses the total particle number $N = 10^4$, a condensate fraction of $N_{\bm 0} / N \approx 0.9$ and the maximal interaction strength $\tilde g(t_\text{end}) = 10^{-5}$.}
	\label{Fig:E_crossover}
\end{figure}
As the effective 1D interaction \eqref{eq:g1D} increases along the dimensional crossover, the deviations observed in the previous sections become significantly large and both the perturbative and the variational calculation are no longer valid in the case $\lambda\gg1$. Figure \ref{Fig:E_crossover} illustrates this finding. In figure \ref{Fig:E_crossover} a) the new eigenenergies are plotted with respect to the non-interacting ground-state energy $E_0 = \hbar\Omega_x(1+\lambda^2)/2$ at the dimensional crossover for a fixed interaction strength $\tilde g(t)$, revealing an increasing effective 1D interaction strength $\tilde g_\text{1D}(t; \lambda)$. Due to the enhanced effective interaction strength, increasing the trap anisotropy shifts the eigenenergies to larger values. Here too, the ground-state energy undergoes the largest change, which is clearly non-linear. Figure \ref{Fig:E_crossover} b) shows that the energy differences likewise increase with the trap-aspect ratio pointing towards the predicted increased photon-photon interaction at the dimensional crossover, cf., reference \cite{Stein2022b}. Thus, the ED method also allows determining the effective photon-photon interaction strength along the dimensional crossover in the same way by spectroscopic measurements as outlined above. For instance, one approach relies on interfering different cavity modes and observing the resulting beating signal.\\
\ref{App:1D} derives an analytic approximation for the eigenvalues, which are numerically calculated by the ED in the deep condensate limit. This approach considers the first four eigenmodes and diagonalises the corresponding 1D Hamiltonian matrix. This reveals the non-linear character of the eigenvalues, shown in figure \ref{Fig:E_crossover} a) to stem from avoided crossings with the next-higher eigenmodes with the same symmetry. Due to the same reason, this method only approximates the ground and first excited state well, whereas the ED and the analytical approximation for the higher excited states agree less for larger anisotropies. This can be avoided by taking into account even higher excited states, which is necessary for going to larger trap anisotropies respectively larger interaction strength, but makes the analytical diagonalisation more difficult. However, if the effective photon-photon interaction strength is small enough for a linearisation, the analytical approximation for the ground and first-excited state represents a result for determining it according to
\begin{align}
    \tilde g_\text{1D}(\lambda) \approx \frac{2\sqrt{2\pi}}{N_{\bm 0}} \left(1-\frac{\Delta E_{1,0}(t)}{\hbar\Omega_x}\right).
\end{align}

\subsection{Width}
\begin{figure}
	\centering
	\includegraphics[width=\linewidth]{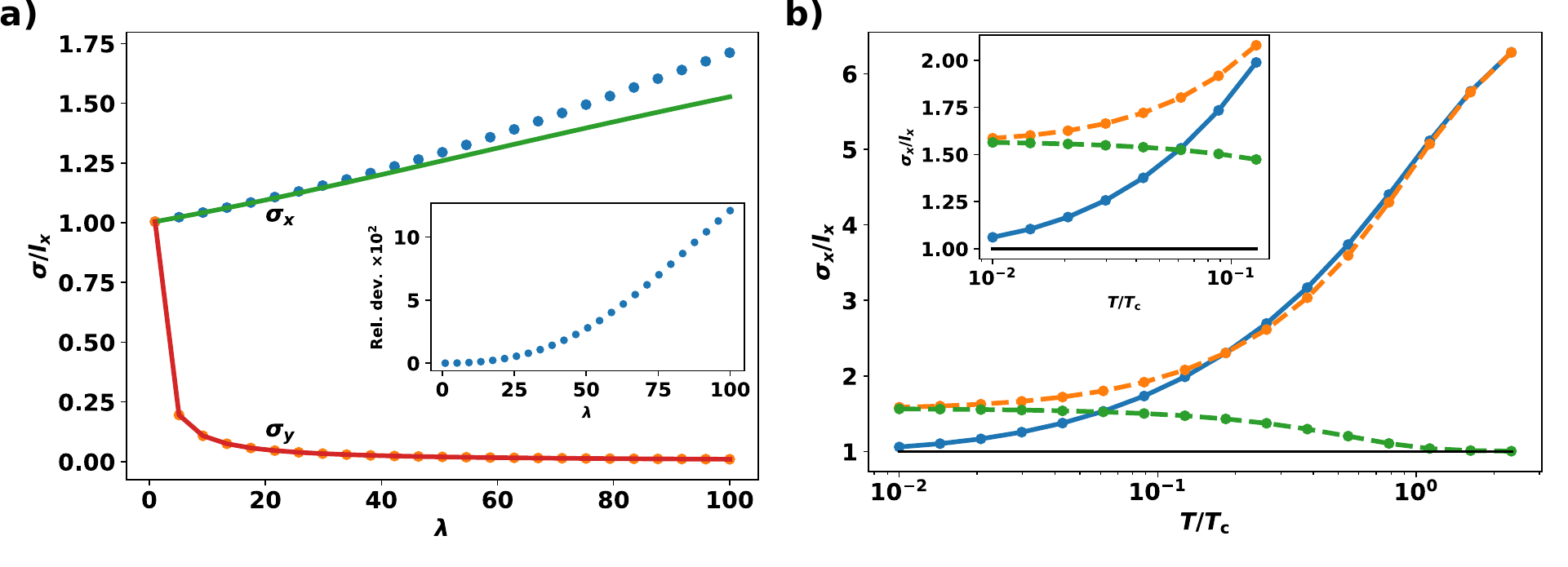}
	\caption{Dimensional crossover. Both plots are for a total particle number $N = 10^4$ and a maximal interaction strength $\tilde g(t_\text{end}) = 10^{-5}$. \textbf{a)} Comparison of condensate width \eqref{eq:total_width} at the end of the experiment for different trap-aspect ratios at fixed interaction strength. Here, the condensate fraction amounts to $N_{\bm 0} / N \approx 0.9$. Dots: results from ED; lines: variational approach. \textbf{b)} Total width of the photon gas at the beginning of the experiment (blue) and at the end of the experiment (orange), as well as the width of the ground state with interaction (green) for trap-aspect ratio $\lambda = 10$. The black line indicates the oscillator length, which is also the width of the ground state without interaction. The drawn lines are a guide to the eye. }
	\label{Fig:Crossover}
\end{figure}
Figure \ref{Fig:Crossover} a) compares the widths from the ED with the results from the variational approach for a fixed interaction strength. Whereas in the squeezed $y$-direction the results from both methods agree quite well at the whole crossover, the results coincide only for small trap-aspect ratios in the un-squeezed $x$-direction. At larger trap-aspect ratios, the results deviate up to several percent, which is caused by the coupling to the thermal cloud. Since the effective photon-photon interaction increases along the dimensional crossover, in the sense of equation \eqref{eq:g1D}, the thermal cloud becomes more and more important in this scenario. Consequently, the ED method plays a crucial role for measuring the effective photon-photon interaction at the dimensional crossover.\\
Also, for larger trap anisotropies the total width of the photon gas can be calculated from \eqref{eq:total_width}, which figure \ref{Fig:Crossover} b) pictures. Here, the blue line depicts again the total width of the photon gas at the beginning of the experiment for a certain thermodynamic temperature $T$. As the experiment runs, the effective photon-photon interaction increases and, thus, also the photon gas width grows during the experiment. Alas, only in the condensed regime this growth is significant and due to the trap anisotropy enhanced in comparison to the situation in the isotropic trap, see figure \ref{Fig:width_temp}. In the thermal regime, however, still no impact of the effective photon-photon interaction is observable.
\section{Summary}
\label{sec:Summary}
The findings presented in this work are crucial for precisely quantifying the effective photon-photon interaction strength in current and future phBEC experiments. The present paper analyses in detail the theory of reference \cite{Stein2022c} by using exact diagonalisation for a harmonically trapped photon gas. The aim consists in determining the impact of the thermo-optic interaction upon the cavity modes. The accurate prediction of the shifts of the eigenenergies allows for a precise interferometric measurement of the emerging photon-photon interaction. The method reproduces formerly derived results and extends these systematically by taking the thermal cloud into account. 
The influence of thermal cloud turns out to be crucial for understanding the dimensional crossover and prevents a possible overestimation of the effective photon-photon interaction strength. Where possible, analytic estimates provide aid for determining the effective photon-photon interaction strength from measurements.

\ack
We thank Antun Bala\v{z}, Georg von Freymann, Milan Radonji\'c, Julian Schulz, Kirankumar Karkihalli Umesh, and Frank Vewinger for insightful discussions. ES and AP acknowledge financial support by the Deutsche Forschungsgemeinschaft (DFG, German Research Foundation) via the Collaborative Research Center SFB/TR185 (Project No. 277625399).
\appendix
\section{Variational Approach}
\label{App:Variational}
This appendix follows reference \cite{Dung2017} and calculates a variational solution for the Hamiltonian \eqref{eq:Ham} at $T=0$ with the potential \eqref{eq:Vho}. In this case, a suitable ansatz for the eigenfunction is given by
\begin{align}
	\psi(\vec x, t) = \sqrt{\frac{\lambda N}{\pi l^2\alpha_x(t)\alpha_y(t)}}\exp\left\{ -\frac{1}{2l^2} \left[ \frac{x^2}{\alpha_x(t)^2} + \frac{y^2}{\alpha_y(t)^2} \right] \right\},		
\end{align}
where $l = \sqrt{\hbar/(m\Omega_x)}$ denotes the oscillator length and $\alpha_x(t), \alpha_y(t)$ stand for the dimensionless variational parameters with the adiabatic time dependency. Consequently, the initial density is given by $n(\vec x, 0) = |\psi(\vec x, 0)|^2$, with $\alpha_x(0) = 1 = \alpha_y(0)$. The standard procedure yields the variational equations 
\begin{align}
	\alpha_x^4 = 1 + \frac{2\tilde g(t) \lambda N}{\pi}\frac{\alpha_x^4}{\sqrt{\left(1+\alpha_x^2\right)^3\left(1+\alpha_y^2\right)}}		
\end{align}
and
\begin{align}
	\alpha_y^4 = 1 + \frac{2\tilde g(t) N}{\pi\lambda}\frac{\alpha_y^4}{\sqrt{\left(1+\alpha_x^2\right)\left(1+\alpha_y^2\right)^3}}\,.
\end{align}
\section{Analytical Approximation in the 1D Case}\label{App:1D}
The aim of this appendix is to work out an analytical approximation for the quasi-1D case of a harmonic potential. Since the trap-aspect ratio determines the interaction strength $\tilde g_\text{1D}(\lambda)$, the latter can reach comparatively large values. Therefore, the first four eigenstates are taken into account and the resulting 1D Hamiltonian matrix is diagonalised. Moreover, the deep condensate limit $N_{\bm 0}\approx N$ is assumed. With the harmonic-oscillator eigenfunctions, the Hamiltonian matrix reads
\begin{align}
    \mathcal H_\text{1D} = \hbar\Omega_x\begin{pmatrix}
    \frac{N_{\bm 0}\tilde g_\text{1D}(\lambda)}{\sqrt{2\pi}} & 0 & -\frac{N_{\bm 0}\tilde g_\text{1D}(\lambda)}{4\sqrt{\pi}} & 0 \\
    0 & 1 + \frac{N_{\bm 0}\tilde g_\text{1D}(\lambda)}{2\sqrt{2\pi}} & 0 & -\frac{N_{\bm 0}\tilde g_\text{1D}(\lambda)\sqrt{3}}{8\sqrt{\pi}}\\
    -\frac{N_{\bm 0}\tilde g_\text{1D}(\lambda)}{4\sqrt{\pi}} & 0 & 2 + \frac{3N_{\bm 0}\tilde g_\text{1D}(\lambda)}{8\sqrt{2\pi}} & 0\\
    0 & -\frac{N_{\bm 0}\tilde g_\text{1D}(\lambda)\sqrt{3}}{8\sqrt{\pi}} & 0 & 3 + \frac{5N_{\bm 0}\tilde g_\text{1D}(\lambda)}{16\sqrt{2\pi}}
    \end{pmatrix},
\end{align}
where the shift due to the unperturbed ground-state energy has been dropped. The Hamiltonian matrix possesses the eigenvalues
\begin{subequations}\label{eq:E1D}
    \begin{align}
        E_0 &= \hbar\Omega_x \left(1 + \frac{11 N_{\bm 0}\tilde g_\text{1D}(\lambda)}{16 \sqrt{2 \pi }} - \sqrt{ 1 -\frac{5 N_{\bm 0}\tilde g_\text{1D}(\lambda)}{8 \sqrt{2 \pi }} + \frac{57 (N_{\bm 0}\tilde g_\text{1D}(\lambda))^2}{512 \pi }}\,\right)\,,\\
        E_1 & =\hbar\Omega_x \left( 2 + \frac{13 N_{\bm 0}\tilde g_\text{1D}(\lambda)}{32 \sqrt{2 \pi }} - \sqrt{ 1 -\frac{3 N_{\bm 0}\tilde g_\text{1D}(\lambda)}{16 \sqrt{2 \pi }} + \frac{105 (N_{\bm 0}\tilde g_\text{1D}(\lambda))^2}{2048 \pi }}\,\right)\,,\\
        E_2 &=\hbar\Omega_x \left( 1 + \frac{11 N_{\bm 0}\tilde g_\text{1D}(\lambda)}{16 \sqrt{2 \pi }} + \sqrt{ 1 -\frac{5 N_{\bm 0}\tilde g_\text{1D}(\lambda)}{8 \sqrt{2 \pi }} + \frac{57 (N_{\bm 0}\tilde g_\text{1D}(\lambda))^2}{512 \pi }}\,\right)\,,\\
        E_3 & = \hbar\Omega_x \left(2 + \frac{13 N_{\bm 0}\tilde g_\text{1D}(\lambda)}{32 \sqrt{2 \pi }} + \sqrt{ 1 -\frac{3 N_{\bm 0}\tilde g_\text{1D}(\lambda)}{16 \sqrt{2 \pi }} + \frac{105 (N_{\bm 0}\tilde g_\text{1D}(\lambda))^2}{2048 \pi }}\,\right)\,.
    \end{align}
\end{subequations}

\section*{References}
\bibliographystyle{unsrt}
\bibliography{refs}

\end{document}